\PassOptionsToPackage{svgnames}{xcolor}
\documentclass[twocolumn]{aastex631}
\usepackage{CJK}

\definecolor{linkcolor}{cmyk}{1,1,0,0}

\usepackage{color,soul}
\usepackage{IEEEtrantools}
\usepackage{scalerel}
\usepackage{dashrule}
\usepackage{totcount}
\newtotcounter{citenum}
\def\oldcite{}
\let\oldcite=\bibcite
\def\bibcite{\stepcounter{citenum}\oldcite}

\defcitealias{RC3}{RC3}
\defcitealias{Baldwin:1981}{BPT}
\defcitealias{Sheth:2010}{S$^4$G}

\definecolor{MATLAB_blue}{rgb}{0,0.4470,0.7410}
\definecolor{MATLAB_red}{rgb}{0.8500,0.3250,0.0980}
\definecolor{MATLAB_orange}{rgb}{0.9290,0.6940,0.1250}
\definecolor{MATLAB_purple}{rgb}{0.4940,0.1840,0.5560}
\definecolor{MATLAB_green}{rgb}{0.4660,0.6740,0.1880}
\definecolor{MATLAB_cyan}{rgb}{0.3010,0.7450,0.9330}
\definecolor{MATLAB_maroon}{rgb}{0.6350,0.0780,0.1840}
\definecolor{LimeGreen}{rgb}{0.1961,0.8039,0.1961}
\definecolor{Orange}{rgb}{1,0.6471,0}

\received{July 11, 2023}
\revised{September 14, 2023}
\accepted{September 15, 2023}
\submitjournal{\textit{The Astrophysical Journal Letters}}

\shorttitle{A Planar Black Hole Mass Scaling Relation}
\shortauthors{Davis \& Jin}

\begin{document}
\begin{CJK*}{UTF8}{gbsn}

\title{Discovery of a Planar Black Hole Mass Scaling Relation for Spiral Galaxies}

\correspondingauthor{Benjamin L.\ Davis}
\email{ben.davis@nyu.edu}

\author[0000-0002-4306-5950]{Benjamin L.\ Davis}
\affil{Center for Astrophysics and Space Science (CASS), New York University Abu Dhabi, PO Box 129188, Abu Dhabi, UAE}

\author[0009-0000-2506-6645]{Zehao Jin (金泽灏)}
\affil{Center for Astrophysics and Space Science (CASS), New York University Abu Dhabi, PO Box 129188, Abu Dhabi, UAE}

\begin{abstract}

Supermassive black holes (SMBHs) are tiny in comparison to the galaxies they inhabit, yet they manage to influence and coevolve along with their hosts.
Evidence of this mutual development is observed in the structure and dynamics of galaxies and their correlations with black hole mass ($M_\bullet$).
For our study, we focus on relative parameters that are unique to only disk galaxies.
As such, we quantify the structure of spiral galaxies via their logarithmic spiral-arm pitch angles ($\phi$) and their dynamics through the maximum rotational velocities of their galactic disks ($v_\mathrm{max}$).
In the past, we have studied black hole mass scaling relations between $M_\bullet$ and $\phi$ or $v_\mathrm{max}$, separately.
Now, we combine the three parameters into a trivariate $M_\bullet$--$\phi$--$v_\mathrm{max}$ relationship that yields best-in-class accuracy in prediction of black hole masses in spiral galaxies.
Because most black hole mass scaling relations have been created from samples of the largest SMBHs within the most massive galaxies, they lack certainty when extrapolated to low-mass spiral galaxies.
Thus, it is difficult to confidently use existing scaling relations when trying to identify galaxies that might harbor the elusive class of intermediate-mass black holes (IMBHs).
Therefore, we offer our novel relationship as an ideal predictor to search for IMBHs and probe the low-mass end of the black hole mass function by utilizing spiral galaxies.
Already with rotational velocities widely available for a large population of galaxies and pitch angles readily measurable from uncalibrated images, we expect that the $M_\bullet$--$\phi$--$v_\mathrm{max}$ fundamental plane will be a useful tool for estimating black hole masses, even at high redshifts.

\end{abstract}

\keywords{
\href{http://astrothesaurus.org/uat/1882}{Astrostatistics (1882)} ---
\href{http://astrothesaurus.org/uat/594}{Galaxy evolution (594)} ---
\href{http://astrothesaurus.org/uat/757}{Hubble classification scheme (757)} ---
\href{http://astrothesaurus.org/uat/816}{Intermediate-mass black holes (816)} ---
\href{http://astrothesaurus.org/uat/907}{Late-type galaxies (907)} ---
\href{http://astrothesaurus.org/uat/1914}{Regression (1914)} ---
\href{http://astrothesaurus.org/uat/2031}{Scaling relations (2031)} ---
\href{http://astrothesaurus.org/uat/1560}{Spiral galaxies (1560)} ---
\href{http://astrothesaurus.org/uat/1561}{Spiral pitch angle (1561)}
}

\section{Introduction}\label{sec:intro}
\end{CJK*}

Black hole mass scaling relations (\textit{i.e.}, relations with central black hole mass as the dependent variable and some physical property of its host galaxy as the independent variable) have evolved and proliferated over the past quarter-century, beginning with the identification of a correlation between black hole mass ($M_\bullet$) and the stellar mass of its host galaxy's bulge \citep{Magorrian:1998}.
The most reliable of these scaling relations are those that are built and calibrated upon samples of galaxies with dynamically-measured black hole masses \citep[\textit{e.g.},][]{Graham:2013,Graham:2023a,Graham:2023c,Graham:2023b,Graham:2023d,Savorgnan:2013,Savorgnan:2016,Savorgnan:2016b,Savorgnan:Thesis,Davis:2017,Davis:2018,Davis:2019c,Davis:2019b,Davis:2019,Davis:2021,Sahu:2019b,Sahu:2019,Sahu:2020,NSahu:2022,NSahu:2022b,Sahu:Thesis,Sahu:2022c,Jin:2023}.\footnote{
We note that alleged biases purported to exist between galaxies that host directly-measured black holes and galaxies from the general population \citep{Shankar:2016} have been quashed \citep{Sahu:2023}.
}
To-date, only about 150 such supermassive black holes (SMBHs) with $M_\bullet\gtrsim10^6\,\mathrm{M}_\sun$ have been directly measured in just the nearest and most massive galaxies.\footnote{See \citet{Jin:2023} for an online database of the sample.}
With SMBHs expected to reside in the hearts of most every massive galaxy \citep{Rees:1984}, one can take their pick of scaling relations \citep[see][for informative reviews]{Graham:2016,D'Onofrio:2021} to perform black hole mass estimates for large numbers of galaxies in surveys to construct black hole mass functions \citep[\textit{e.g.},][]{Graham:2007,Davis:2014,Mutlu-Pakdil:2016}.

The accuracy of scaling relations can vary significantly based on which independent variable is selected; some variables are less accurate or not applicable for certain galaxy morphologies (\textit{e.g.}, bulge relations are useless for bulgeless galaxies).
Moreover, the prevalence of morphologically-dependent black hole mass scaling relations \citep[\textit{e.g.},][]{Davis:2018,Davis:2019b,Sahu:2019b} hints that particular independent variables alone might not be sufficient to cover different galaxy morphologies.
Specifically, this is evident from the different coefficients required for the same variables when applied to separate morphologies in isolation.
As such, it is problematic to be restricted to using only one predictor of black hole mass.
Thus, we seek a methodology that incorporates multiple mass predictors in one relation.

Dynamical measurements of smaller black holes like intermediate-mass black holes (IMBHs) are much more difficult to obtain because the gravitational sphere of influence radius of a black hole is directly proportional to the mass of the black hole.
As expected, an observational bias exists among the catalog of directly-measured black holes, \textit{i.e.}, only black holes that are sufficiently massive and/or nearby are measurable \citep{Batcheldor:2010}.
Our current sample of 145 dynamically-measured black holes ranges from $4\times10^5\,\mathrm{M}_\sun\lesssim M_\bullet\lesssim2\times10^{10}\,\mathrm{M}_\sun$ \citep{Jin:2023}.
Although this sample reaches down almost to the IMBH regime ($10^3\leq M_\bullet<10^5\,\mathrm{M}_\sun$), the sample of black holes is very top heavy with a median mass of $\approx$$10^8\,\mathrm{M}_\sun$.
Therefore, interpolation of existing black hole mass scaling relations are incapable of predicting IMBHs and extrapolation is heavily reliant on the largest SMBHs.
As such, several studies have instead relied on meta-analyses to combine the predictions of multiple black hole mass scaling relations to more securely extrapolate down into the IMBH regime \citep{Koliopanos:2017,Graham:Soria:2019,Graham:2019,Davis:2020,Davis:2023b}.

In a larger study \citep{Jin:2023}, we used modern machine learning methods to identify higher-dimensional ($n$-D) black hole mass scaling relations that have lower intrinsic scatters than existing two-dimensional (2-D) black hole mass scaling relations.
With 145 galaxies and as much as a hundred different measured quantities for every galaxy, the task of checking the vast number of permutations of possible $n$-D black hole mass scaling relations is an immense undertaking.
Therefore, we applied modern machine learning methods to find the best scaling relations, which ideally are an optimized combination of accuracy and simplicity.
For this task, we ran symbolic regression software \href{https://github.com/MilesCranmer/PySR/tree/v0.12.3}{\textcolor{linkcolor}{\texttt{PySR}}} \citep{Cranmer:2023} to find the best combination of variables and mathematical operations to describe our dataset of directly-measured SMBH masses and their host galaxy parameters.

In this letter, we describe in detail one such solution found by of our study: a trivariate relationship between $M_\bullet$, logarithmic spiral-arm pitch angle ($\phi$), and maximum rotational velocity ($v_\mathrm{max}$).
We will present our sample, fit, and analysis of the planar black hole mass scaling relation in \S\ref{sec:FP}.
In \S\ref{sec:discussion}, we will discuss benefits, reasons, comparisons, implications, and utility of our $M_\bullet$--$\phi$--$v_\mathrm{max}$ fundamental plane.
Finally, we provide a summary of our findings and remark on future work (in \S\ref{sec:conclusions}).
We represent black hole masses ($M_\bullet$) throughout this work as logarithmic (solar) masses ($\mathcal{M}_\bullet$), such that $\mathcal{M}_\bullet\equiv\log(M_\bullet/\mathrm{M}_\sun)$.
All uncertainties are quoted at $1\,\sigma \equiv 68.3\%$ confidence intervals; median absolute deviations are given as uncertainties associated with medians.

\section{Data and Analysis}\label{sec:FP}
 
\subsection{Sample}

Our sample consists of all spiral galaxies with $M_\bullet$, $\phi$, and $v_\mathrm{max}$ measurements from \citet{Davis:2019}.
This yields a set of 41 galaxies (not including the Milky Way), all with dynamically-measured black hole masses \citep[see references compiled by][]{Davis:2017,Davis:2019b}.
Pitch angles were consistently measured by \citet{Davis:2017,Davis:2019}\footnote{For details regarding the measurement of galactic logarithmic spiral-arm pitch angles, see additional reading \citep{Davis:2012,sparcfire,Davis:Thesis,Shields:2022}.} and rotational velocities compiled by \citet[][see references therein]{Davis:2019}.
This sample that we use to construct the planar relation is listed in Table~\ref{tab:sample}.

\startlongtable
\begin{deluxetable}{lrll}
\tablecolumns{4}
\tablecaption{Sample of Spiral Galaxies}\label{tab:sample}
\tablehead{
\colhead{Galaxy} & \colhead{$|\phi|$} & \colhead{$v_\mathrm{max}$} & \colhead{$\mathcal{M}_\bullet$} \\
\colhead{} & \colhead{[$\degr$]} & \colhead{$\left[\frac{\text{km}}{{\text{s}}}\right]$} & \colhead{[dex]} \\
\colhead{(1)} & \colhead{(2)} & \colhead{(3)} & \colhead{(4)}
}
\startdata
\object{Circinus} & $17\fdg0\pm3\fdg9$ & $153\pm7$ & $6.25\pm0.11$ \\
\object{IC~2560} & $22\fdg4\pm1\fdg7$ & $196\pm3$ & $6.52\pm0.11$ \\
\object{Milky~Way}\tablenotemark{$^\ast$} & $13\fdg1\pm0\fdg6$ & $198\pm6$ & $6.60\pm0.02$ \\
\object{NGC~224} & $8\fdg5\pm1\fdg3$ & $257\pm6$ & $8.15\pm0.16$ \\
\object{NGC~253} & $13\fdg8\pm2\fdg3$ & $196\pm3$ & $7.00\pm0.30$ \\
\object{NGC~613} & $15\fdg8\pm4\fdg3$ & $289\pm5$ & $7.57\pm0.15$ \\
\object{NGC~1068} & $17\fdg3\pm1\fdg9$ & $192\pm12$ & $6.75\pm0.08$ \\
\object{NGC~1097} & $9\fdg5\pm1\fdg3$ & $241\pm34$ & $8.38\pm0.04$ \\
\object{NGC~1300} & $12\fdg7\pm2\fdg0$ & $189\pm28$ & $7.86\pm0.14$ \\
\object{NGC~1320} & $19\fdg3\pm2\fdg0$ & $183\pm13$ & $6.77\pm0.22$ \\
\object{NGC~1365} & $11\fdg4\pm0\fdg1$ & $198\pm3$ & $6.60\pm0.30$ \\
\object{NGC~1398} & $9\fdg7\pm0\fdg7$ & $289\pm7$ & $8.03\pm0.11$ \\
\object{NGC~1566} & $17\fdg8\pm3\fdg7$ & $154\pm14$ & $6.83\pm0.30$ \\
\object{NGC~1672} & $15\fdg4\pm3\fdg6$ & $213\pm8$ & $7.70\pm0.10$ \\
\object{NGC~2273} & $15\fdg2\pm3\fdg9$ & $211\pm16$ & $6.95\pm0.06$ \\
\object{NGC~2748} & $6\fdg8\pm2\fdg2$ & $188\pm27$ & $7.54\pm0.21$ \\
\object{NGC~2960} & $14\fdg9\pm1\fdg9$ & $257\pm34$ & $7.07\pm0.05$ \\
\object{NGC~2974} & $10\fdg5\pm2\fdg9$ & $284\pm26$ & $8.23\pm0.07$ \\
\object{NGC~3031} & $13\fdg4\pm2\fdg3$ & $237\pm10$ & $7.83\pm0.09$ \\
\object{NGC~3079} & $20\fdg6\pm3\fdg8$ & $216\pm6$ & $6.38\pm0.12$ \\
\object{NGC~3227} & $7\fdg7\pm1\fdg4$ & $240\pm10$ & $7.97\pm0.14$ \\
\object{NGC~3368} & $14\fdg0\pm1\fdg4$ & $218\pm15$ & $6.89\pm0.11$ \\
\object{NGC~3393} & $13\fdg1\pm2\fdg5$ & $193\pm48$ & $7.49\pm0.05$ \\
\object{NGC~3627} & $18\fdg6\pm2\fdg9$ & $188\pm7$ & $6.94\pm0.09$ \\
\object{NGC~4151} & $11\fdg8\pm1\fdg8$ & $272\pm16$ & $7.69\pm0.37$ \\
\object{NGC~4258} & $13\fdg2\pm2\fdg5$ & $222\pm8$ & $7.60\pm0.01$ \\
\object{NGC~4303} & $14\fdg7\pm0\fdg9$ & $214\pm7$ & $6.78\pm0.17$ \\
\object{NGC~4388} & $18\fdg6\pm2\fdg6$ & $180\pm5$ & $6.90\pm0.10$ \\
\object{NGC~4395} & $22\fdg7\pm3\fdg6$ & $145\pm11$ & $5.62\pm0.17$ \\
\object{NGC~4501} & $12\fdg2\pm3\fdg4$ & $272\pm4$ & $7.31\pm0.08$ \\
\object{NGC~4594} & $5\fdg2\pm0\fdg4$ & $277\pm22$ & $8.81\pm0.03$ \\
\object{NGC~4699} & $5\fdg1\pm0\fdg4$ & $258\pm7$ & $8.27\pm0.09$ \\
\object{NGC~4736} & $15\fdg0\pm2\fdg3$ & $182\pm5$ & $6.83\pm0.11$ \\
\object{NGC~4826} & $24\fdg3\pm1\fdg5$ & $167\pm9$ & $6.18\pm0.12$ \\
\object{NGC~4945} & $22\fdg2\pm3\fdg0$ & $171\pm2$ & $6.13\pm0.30$ \\
\object{NGC~5055} & $4\fdg1\pm0\fdg4$ & $270\pm14$ & $8.94\pm0.10$ \\
\object{NGC~5495} & $13\fdg3\pm1\fdg4$ & $202\pm43$ & $7.04\pm0.08$ \\
\object{NGC~5765b} & $13\fdg5\pm3\fdg9$ & $238\pm15$ & $7.72\pm0.05$ \\
\object{NGC~6926} & $9\fdg1\pm0\fdg7$ & $246\pm10$ & $7.68\pm0.50$ \\
\object{NGC~7582} & $10\fdg9\pm1\fdg6$ & $200\pm9$ & $7.72\pm0.12$ \\
\object{UGC~3789} & $10\fdg4\pm1\fdg9$ & $210\pm14$ & $7.07\pm0.05$ \\
\object{UGC~6093} & $10\fdg2\pm0\fdg9$ & $170\pm59$ & $7.41\pm0.03$ \\
\enddata
\tablecomments{
This sample of 41 spiral galaxies (not including the Milky Way) consists of all spiral galaxies with $\phi$, $v_\mathrm{max}$, and $M_\bullet$ measurements from our larger sample of all galaxy types.
The full parent dataset of 145 galaxies is available online via \citet{Jin:2023}.
\textbf{Column~(1):} galaxy name.
\textbf{Column~(2):} absolute value of the \emph{face-on} (\textit{i.e.}, de-projected from the plane of the sky) spiral-arm pitch angle (in degrees), from \citet{Davis:2017,Davis:2019}.
\textbf{Column~(3):} physical maximum velocity rotation (in km\,s$^{-1}$) corrected for inclination and compiled by \citet{Davis:2019} from references therein.
\textbf{Column~(4):} dynamical black hole mass (in dex, solar masses) measurement compiled by \citet{Davis:2017,Davis:2019b} from references therein.
}
\tablenotetext{\ast}{We do not include the Milky Way in our preferred determination of the fundamental plane (see \S\ref{app:MW} for further details).
}
\end{deluxetable}

\begin{figure*}
\includegraphics[clip=true, trim= 0mm 0mm 0mm 0mm, height=4.65cm]{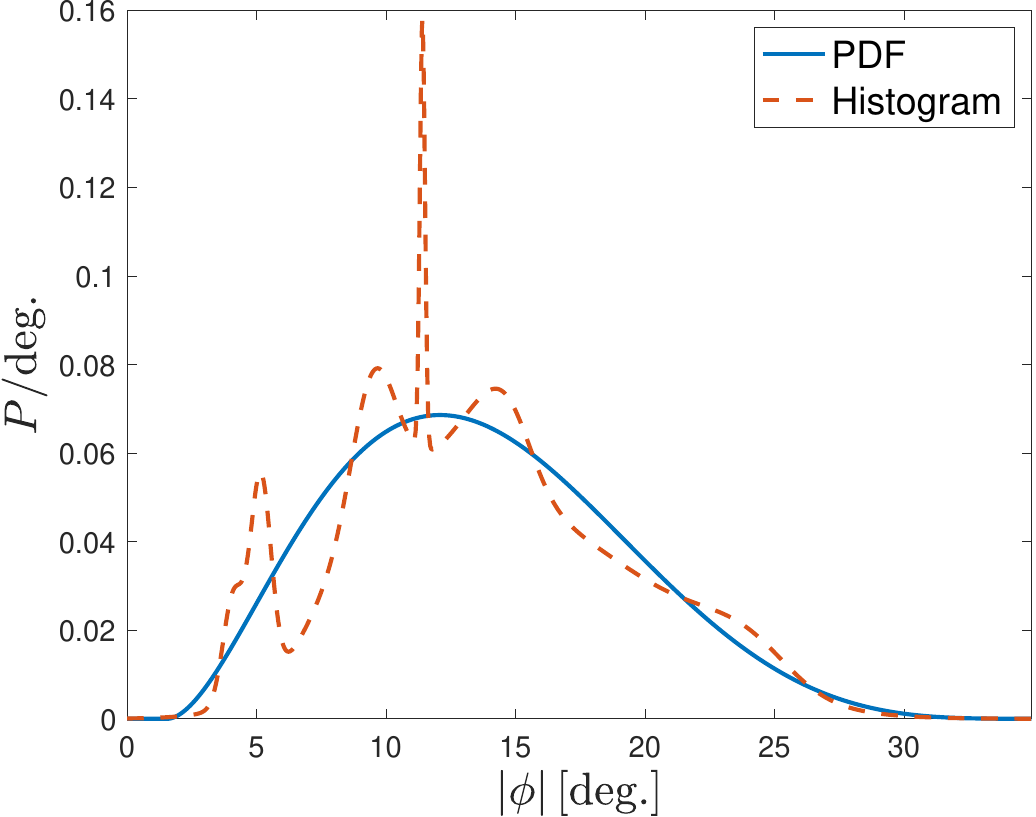}
\includegraphics[clip=true, trim= 0mm 0mm 0mm 0mm, height=4.65cm]{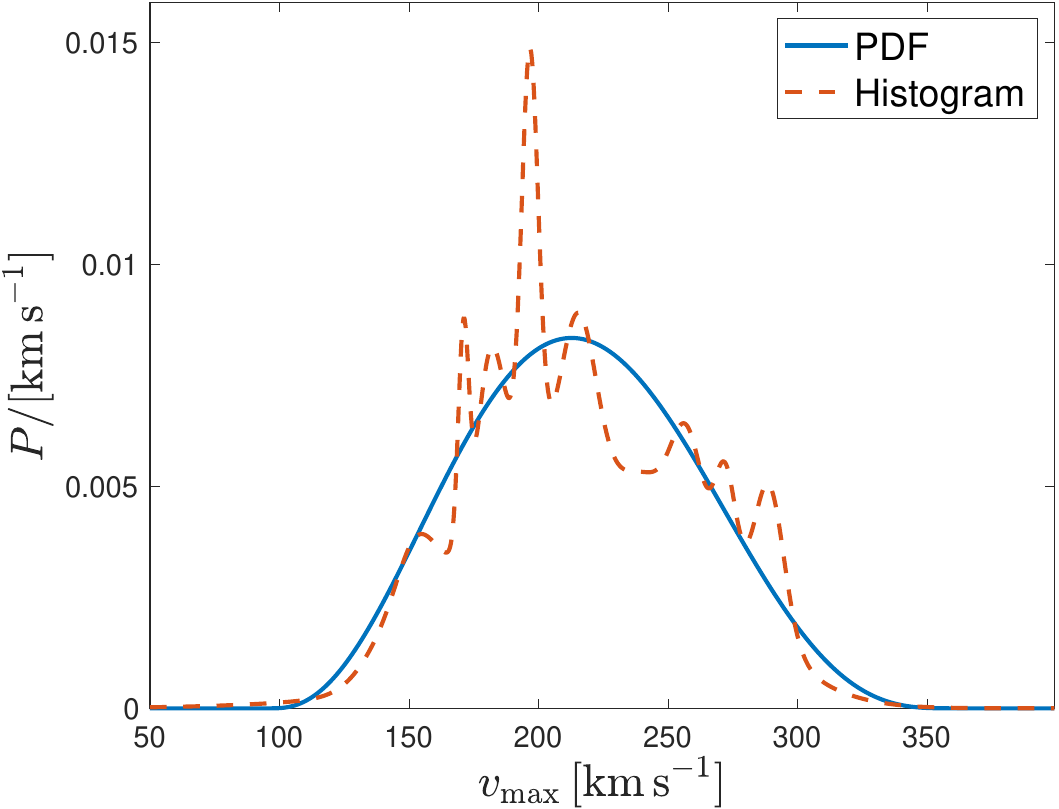}
\includegraphics[clip=true, trim= 0mm 0mm 0mm 0mm, height=4.65cm]{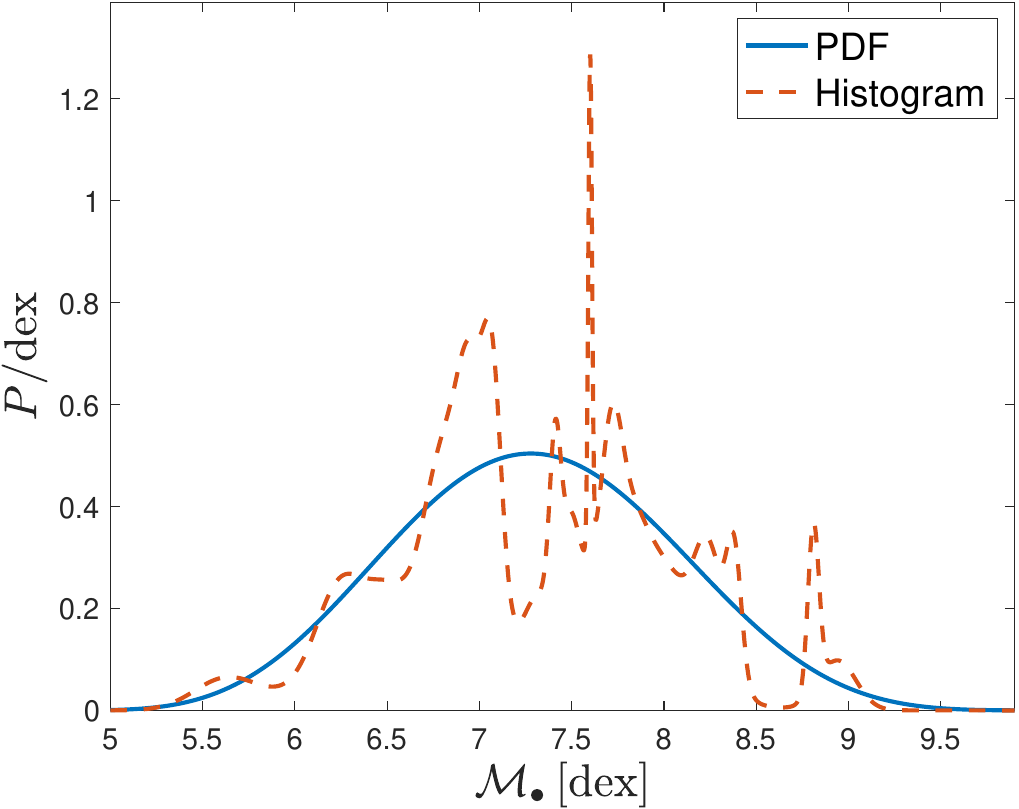}
\caption{
The distributions of $\phi$ (\emph{left}), $v_\mathrm{max}$ (\emph{middle}), and $M_\bullet$ (\emph{right}) for our sample of 41 spiral galaxies.
The smoothed histograms (\textcolor{MATLAB_red}{{\hdashrule[0.35ex]{8mm}{1pt}{1mm}}}) are generated from the summation of all 41 respective measurements with their uncertainties for each galaxy (\textit{i.e.}, kernel density estimations).
The PDFs (\textcolor{MATLAB_blue}{{\hdashrule[0.35ex]{8mm}{0.4mm}{}}}) are skew-kurtotic-normal distributions fit to each histogram.
\textbf{Left}: mean = $13\fdg7\pm5\fdg4$, median = $13\fdg2\pm3\fdg6$, and peak probability at $12\fdg1\pm0\fdg9$.
\textbf{Middle}: mean = $217\pm44\,\mathrm{km\,s}^{-1}$, median = $211\pm32\,\mathrm{km\,s}^{-1}$, and peak probability at $213\pm7\,\mathrm{km\,s}^{-1}$.
\textbf{Right}: mean = $7.30\pm0.75$\,dex, median = $7.27\pm0.51$\,dex, and peak probability at $7.28\pm0.12$\,dex.
}
\label{fig:PDFs}
\end{figure*}

The sample of SMBH host galaxies exhibits a broad range in each of the three variables.
To illustrate this, we have plotted probability density functions (PDFs) of each parameter in Figure~\ref{fig:PDFs}.
From these distributions, we normalize the $\phi$ and $v_\mathrm{max}$ values about their respective medians to minimize the covariance between the estimated coefficients during regression analysis.
Following an initial symbolic regression, we then use the outputs from \href{https://github.com/MilesCranmer/PySR/tree/v0.12.3}{\textcolor{linkcolor}{\texttt{PySR}}} \citep{Cranmer:2023} as our input initial guesses for \href{https://github.com/CullanHowlett/HyperFit}{\textcolor{linkcolor}{\texttt{Hyper-Fit}}} \citep{Robotham:2015,Robotham:2016}.\footnote{
The combination of \href{https://github.com/MilesCranmer/PySR/tree/v0.12.3}{\textcolor{linkcolor}{\texttt{PySR}}} and \href{https://github.com/CullanHowlett/HyperFit}{\textcolor{linkcolor}{\texttt{Hyper-Fit}}} is necessary to produce a relation that considers and takes into account the errors on individual measurements.
\href{https://github.com/MilesCranmer/PySR/tree/v0.12.3}{\textcolor{linkcolor}{\texttt{PySR}}} only considers the uncertainties on the dependent variable (\textit{i.e.}, $M_\bullet$) without accounting for the uncertainties on the independent variables (\textit{i.e.}, $\phi$ and $v_\mathrm{max}$), and produces a best-fit relation without uncertainties on the derived coefficients nor computing the intrinsic scatter of the relation.
Whereas, \href{https://github.com/CullanHowlett/HyperFit}{\textcolor{linkcolor}{\texttt{Hyper-Fit}}} is able to refine the fit found by \href{https://github.com/MilesCranmer/PySR/tree/v0.12.3}{\textcolor{linkcolor}{\texttt{PySR}}} while taking into account errors on every measurement, producing uncertainties on each derived coefficient, and determining the intrinsic scatter of the plane.
}

\subsection{Finding the Plane via Machine Learning}

Symbolic Regression is a sub-field of machine learning that aims to find mathematical expressions that best fit a given set of data.
Symbolic Regression searches over equations made of possible selections and combinations of variables, operators, and constants, and judges these equations with a score defined by both accuracy and simplicity.
In this work, we adopt the symbolic regression package \href{https://github.com/MilesCranmer/PySR/tree/v0.12.3}{\textcolor{linkcolor}{\texttt{PySR}}} \citep{Cranmer:2023}, which conducts the equation search through a multi-population evolutionary algorithm. 
The accuracy is defined by the mean squared error loss, and the simplicity is characterized by a complexity score, where each use of variables, operators, and constants adds some pre-defined complexity.
The final score of an equation aims to maximize the accuracy and penalize the complexity with a parsimony constant.

The variable pool that we input to \href{https://github.com/MilesCranmer/PySR/tree/v0.12.3}{\textcolor{linkcolor}{\texttt{PySR}}} includes all of the data from \citet{Davis:2017,Davis:2019}, parameters modeled by the bulge/disk decompositions of \citet{Davis:2019b}, and their derived spheroid stellar density properties \citep{NSahu:2022b}.
These variables included, but are not limited to: pitch angle, central stellar velocity dispersion, maximum rotational velocity, galaxy stellar mass, and several properties of the spheroid, including S\'ersic index, half-light radius, stellar mass, and densities (apparent, projected, and de-projected).
We also included all available measurements from HyperLeda \citep{Makarov:2014}, \textit{e.g.}, colors, diameters, \textit{etc.}
The variable pool also included multiple copies of each variable in different forms of natural numbers, their logarithms, and trigonometric functions (\textit{e.g.}, $|\phi|$ or $\tan|\phi|$ and $v_\mathrm{max}$ or $\log{v_\mathrm{max}}$).
The arithmetic operator pool is simply $+$, $-$, $\times$, and $\div$, with additional $\log_{10}$, power, and exponentiation in rare cases.
Based upon a search on these criteria, \href{https://github.com/MilesCranmer/PySR/tree/v0.12.3}{\textcolor{linkcolor}{\texttt{PySR}}} found an optimal correlation between $\mathcal{M}_\bullet$, $\tan|\phi|$, and $\log (v_\mathrm{max}/\mathrm{km}\,\mathrm{s}^{-1})$.
A presentation and discussion of other interesting scaling relations found by \href{https://github.com/MilesCranmer/PySR/tree/v0.12.3}{\textcolor{linkcolor}{\texttt{PySR}}} are beyond the intended scope of this letter, they will be addressed in a more comprehensive work \citep{Jin:2023}.

\begin{figure*}
\includegraphics[clip=true, trim= 14mm 52mm 4mm 59mm, width=0.5\textwidth]{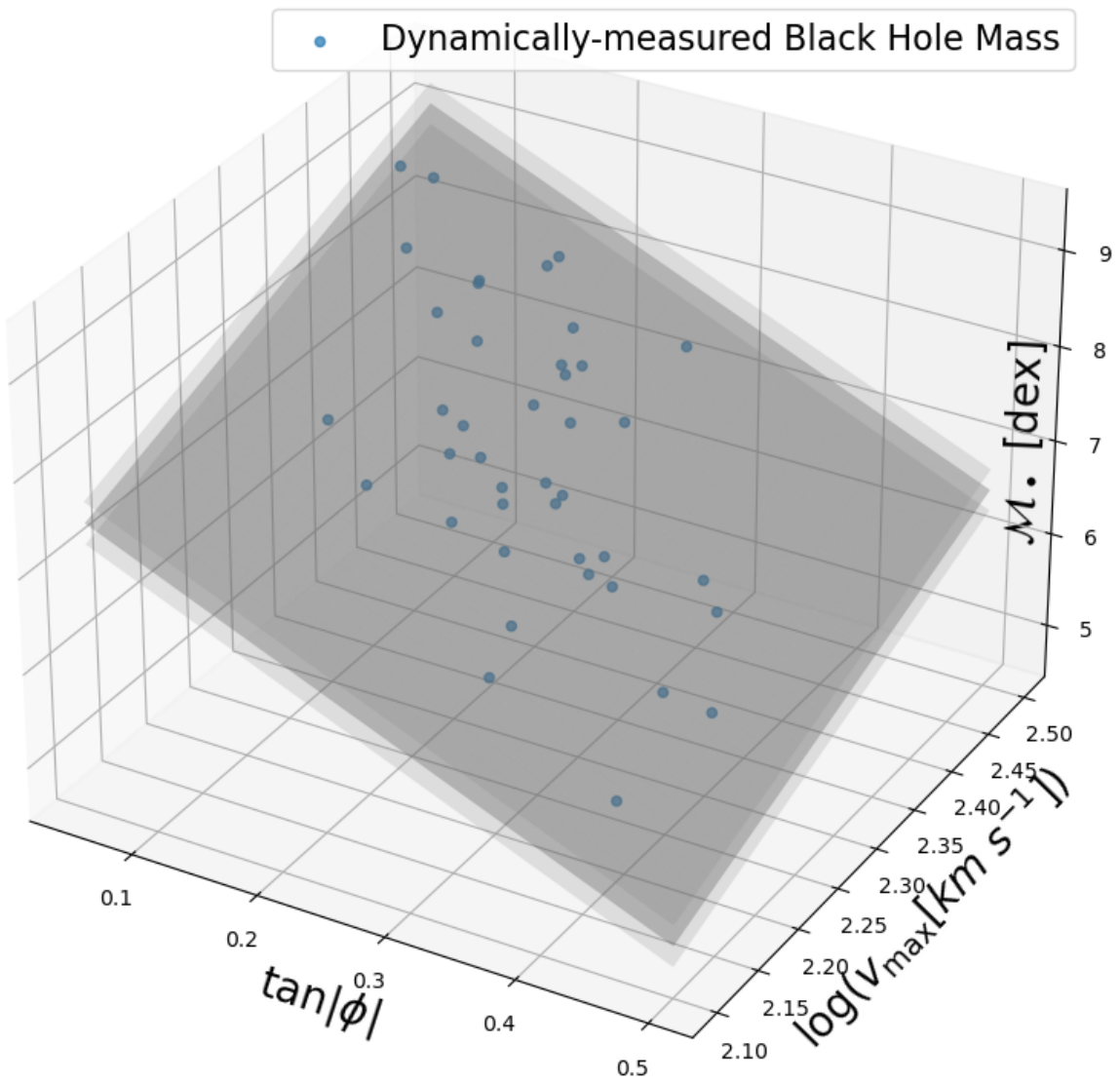}
\includegraphics[clip=true, trim= 56mm 14mm 46mm 16mm, width=0.5\textwidth]{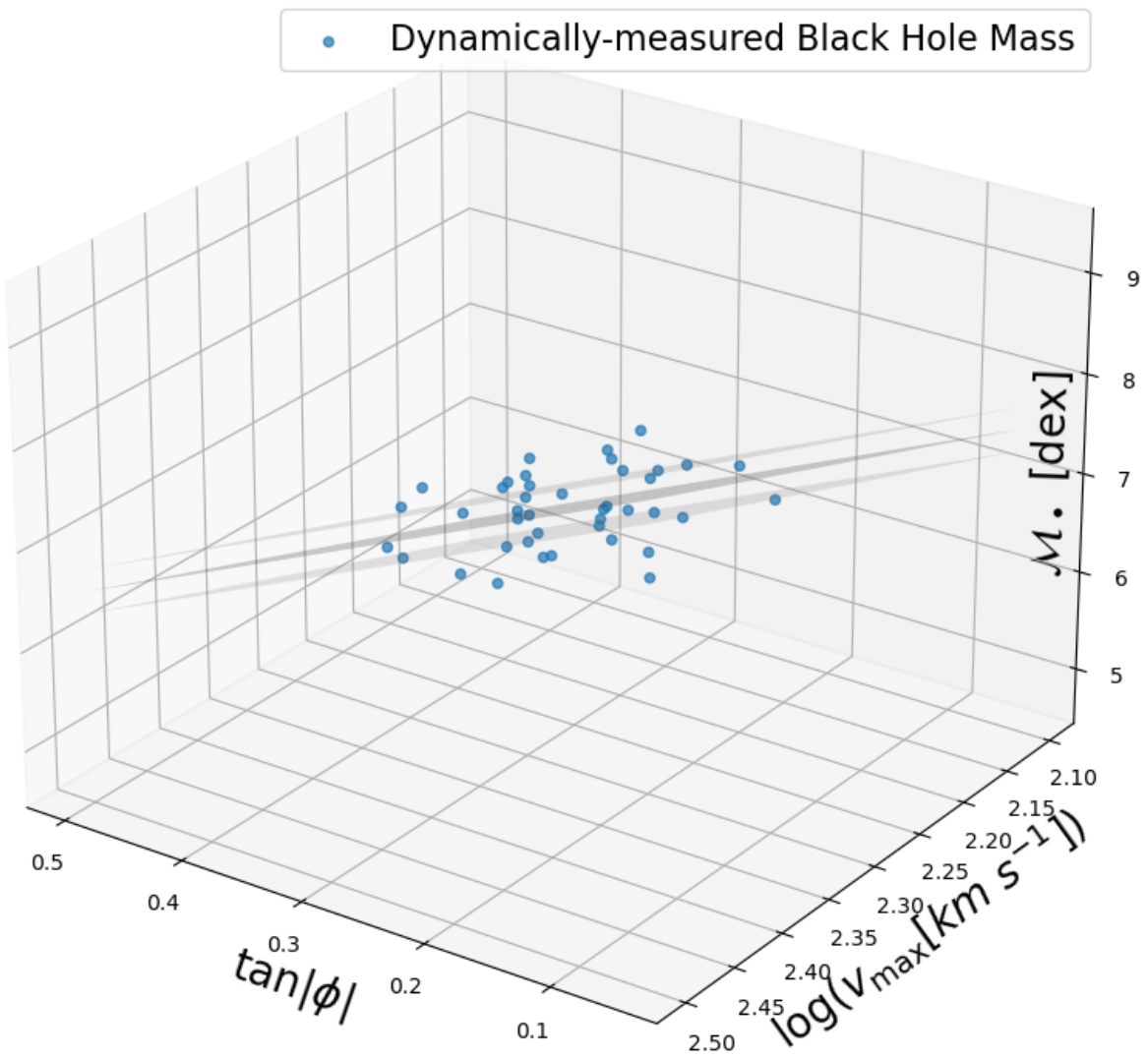}
\includegraphics[clip=true, trim= 37mm 84mm 30mm 93mm, width=0.5\textwidth]{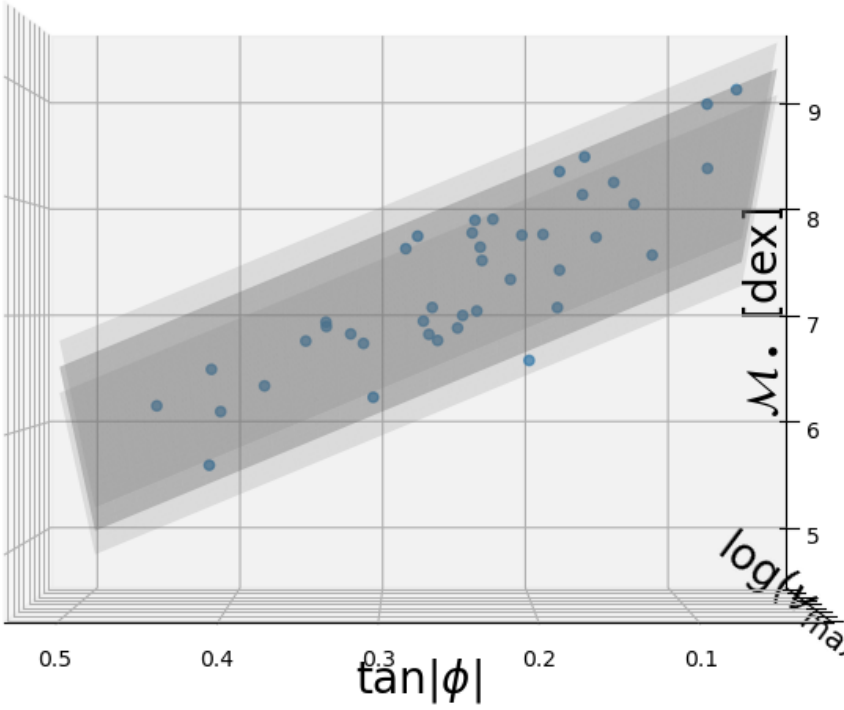}
\includegraphics[clip=true, trim= 30mm 84mm 30mm 93mm, width=0.5\textwidth]{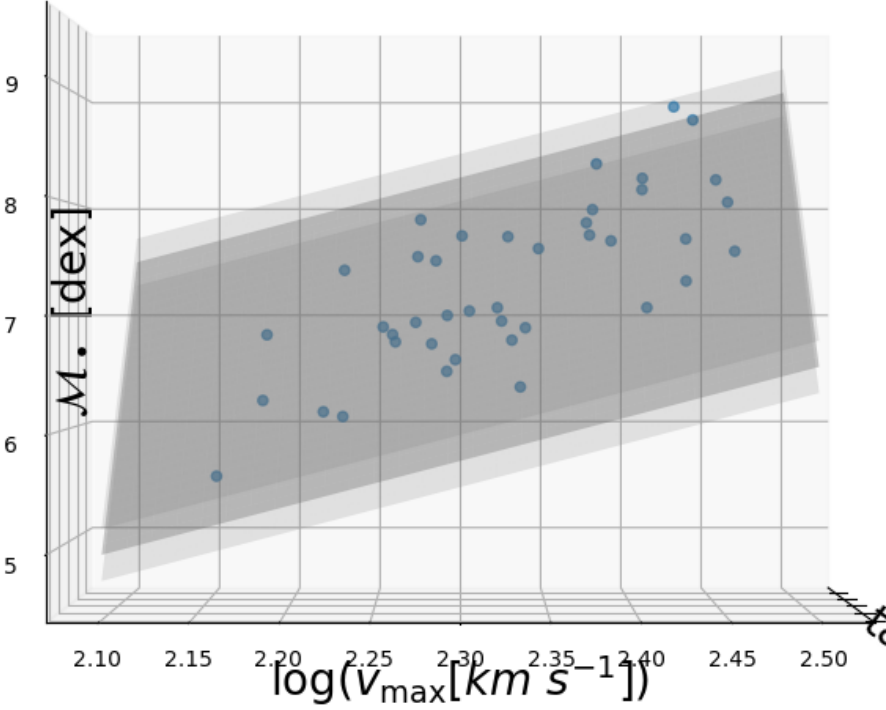}
\caption{
The three-dimensional plot (viewed from four different vantage points) of the planar $M_\bullet$--$\phi$--$v_\mathrm{max}$ relationship (Equation~\ref{eqn:FP}).
Onto the surface (\textcolor{lightgray}{$\blacklozenge$}), we show the locations of the 41 spiral galaxies (\textcolor{MATLAB_blue}{$\bullet$}) from \citet{Jin:2023} used to define the plane.
The fainter gray planes above and below the darker gray middle plane depict the intrinsic scatter bounds ($\pm0.22$\,dex in the $\mathcal{M}_\bullet$-direction).
This plot illustrates that our galaxies are dispersed over the area of the plane, demonstrating a lack of degeneracy between the parameters by the apparent embedding of the two-dimensional manifold (\textit{i.e.}, surface) in three-dimensional space. 
For an animation of this plot (showing also the intrinsic scatter bounds above and below the plane), see the following link, \url{http://surl.li/iggdg}.
}
\label{fig:FP}
\end{figure*}

The functional form and initial parameters were identified by \href{https://github.com/MilesCranmer/PySR/tree/v0.12.3}{\textcolor{linkcolor}{\texttt{PySR}}} and refined via \href{https://github.com/CullanHowlett/HyperFit}{\textcolor{linkcolor}{\texttt{Hyper-Fit}}}.
The final fitted equation for the $M_\bullet$--$\phi$--$v_\mathrm{max}$ relationship is
\begin{IEEEeqnarray}{lCl}
\label{eqn:FP}
\mathcal{M}_\bullet \sim\, &\mathcal{N}&[\mu=\alpha(\tan|\phi|-0.24) \\ \nonumber
&+& \beta\log\left(\frac{v_\mathrm{max}}{211\,\mathrm{km\,s}^{-1}}\right) +\gamma, \\ \nonumber
&\sigma&=0.22\pm0.06],\nonumber
\end{IEEEeqnarray}
with $\alpha=-5.58\pm0.06$, $\beta=3.96\pm0.06$, $\gamma=7.33\pm0.05$, and intrinsic scatter ($\sigma$) in the $\mathcal{M}_\bullet$-direction.\footnote{\href{https://github.com/CullanHowlett/HyperFit}{\textcolor{linkcolor}{\texttt{Hyper-Fit}}} minimizes the intrinsic scatter orthogonal to the plane and then performs a transformation from normal to Cartesian coordinates and outputs the intrinsic scatter along the axis of the dependent variable.}
We present a 3-D plot of the resulting plane in Figure~\ref{fig:FP}.
The orientation of the plane intuitively matches the expectation of the extreme cases:
\begin{itemize}
\item the \textbf{most} massive black holes reside in host galaxies with \underline{tightly} wound spiral arms \emph{and} \underline{high} rotational velocities,
\item the \textbf{least} massive black holes are found in galaxies with \underline{loosely} wound spiral arms \emph{and} \underline{low} rotational velocities,
\item \textbf{no} black holes are found in galaxies with \underline{tightly} wound spiral arms \emph{and} \underline{low} rotational velocities, and
\item \textbf{no} black holes are found in galaxies with \underline{loosely} wound spiral arms \emph{and} \underline{high} rotational velocities.
\end{itemize}
For additional analyses and discussions, see \citet{Jin:2023} for higher-dimensional relations featuring all galaxy types.\footnote{
In the symbolic regression analysis of the spiral galaxies in our sample, we investigated and considered higher-dimensional versions of Equation~\ref{eqn:FP} that incorporated additional quantities such as colors and bulge-to-total ratios.
However, none of our higher-dimensional combinations improved upon the optimization of Equation~\ref{eqn:FP} with the added expense of increased complexity and error propagation.
We find the planar relation is valid because it is built upon parameters ($\phi$ and $v_\mathrm{max}$) that are unique to disk galaxies.
In our forthcoming work \citep{Jin:2023}, we will present higher-dimensional relations that we were able to find due to the larger combined sample of late-type and early-type galaxies with their more varied ranges of colors, bulge-to-total ratios, etc., as compared to our sample of just spiral galaxies in the current work.
}

\subsection{Error Analysis}

One sign of a robust multi-parameter relationship is when different variables contribute equitably.
That is, one variable should not have an overly-dominant influence on the relationship.
Therefore, we need to check the relative change in $M_\bullet$ when there is a proportional change in $\phi$ or $v_\mathrm{max}$.
To check this, we test Equation~\ref{eqn:FP} with equivalent 10\% variations in the median values of $\phi$ or $v_\mathrm{max}$.
Doing so, we find that a change of 10\% in $\phi$ leads to a 37.39\% change in $M_\bullet$ or a 10\% change in $v_\mathrm{max}$ leads to a 48.59\% change in $M_\bullet$.
Ergo, in terms of overall weight, $v_\mathrm{max}$ accounts for a slight majority (56.51\%) of the variation in $M_\bullet$ as compared to a similarly-sized variation in $\phi$.
Thus, neither variable has an outsized influence on $M_\bullet$.

We used the \href{https://github.com/CullanHowlett/HyperFit}{\textcolor{linkcolor}{\texttt{Hyper-Fit}}} routine to robustly fit the equation of the plane to the ($\phi$, $v_\mathrm{max}$, $M_\bullet$) variable set, with consideration of the individual uncertainties on all three parameters and accounting for intrinsic scatter in the relation.
As suggested by its name, \href{https://github.com/CullanHowlett/HyperFit}{\textcolor{linkcolor}{\texttt{Hyper-Fit}}} is uniquely designed to fit ``linear models to multi-dimensional data with multi-variate Gaussian uncertainties.''
Additionally, \href{https://github.com/CullanHowlett/HyperFit}{\textcolor{linkcolor}{\texttt{Hyper-Fit}}} calculates the intrinsic scatter of a scaling relation, which can be considered as the root-mean-square deviation in the observed data from the fitted function in the case of zero measurement error.\footnote{For a more detailed description of intrinsic scatter and its determination in galaxy scaling relations, see \citet{Stone:2021}.}
Therefore, intrinsic scatter is the ideal parameter to judge and compare the accuracy of various scaling relations.

Our determination of a fundamental plane of black hole mass in spiral galaxies is ultimately advantageous because of its combination of the $M_\bullet$--$\phi$ ($\sigma=0.33\pm0.08$\,dex) and $M_\bullet$--$v_\mathrm{max}$ ($\sigma\sim0.45$\,dex) relations, reducing the intrinsic scatter down to $\sigma=0.22\pm0.06$\,dex in the $\mathcal{M}_\bullet$-direction.
Previously, the $M_\bullet$--$\phi$ relation had the lowest level of intrinsic scatter among black hole mass scaling relations for spiral galaxies.
This reduction down to $\sigma=0.22\pm0.06$\,dex (now below a factor of $\log{2}\approx0.3$\,dex) with the planar relation significantly improves upon the previous accuracy of the $M_\bullet$--$\phi$ relation, which was already well below intrinsic scatters available for black hole mass scaling relations built from samples of late- and early-type galaxies.
With such a low-level of intrinsic scatter, this makes the $M_\bullet$--$\phi$--$v_\mathrm{max}$ relation the preeminent scaling relation for black hole mass in spiral galaxies.

\section{Discussion}\label{sec:discussion}

\subsection{The Benefit of Combining Two Relations}

\begin{figure*}
\includegraphics[clip=true, trim=31mm 11mm 49mm 25mm, width=\textwidth]{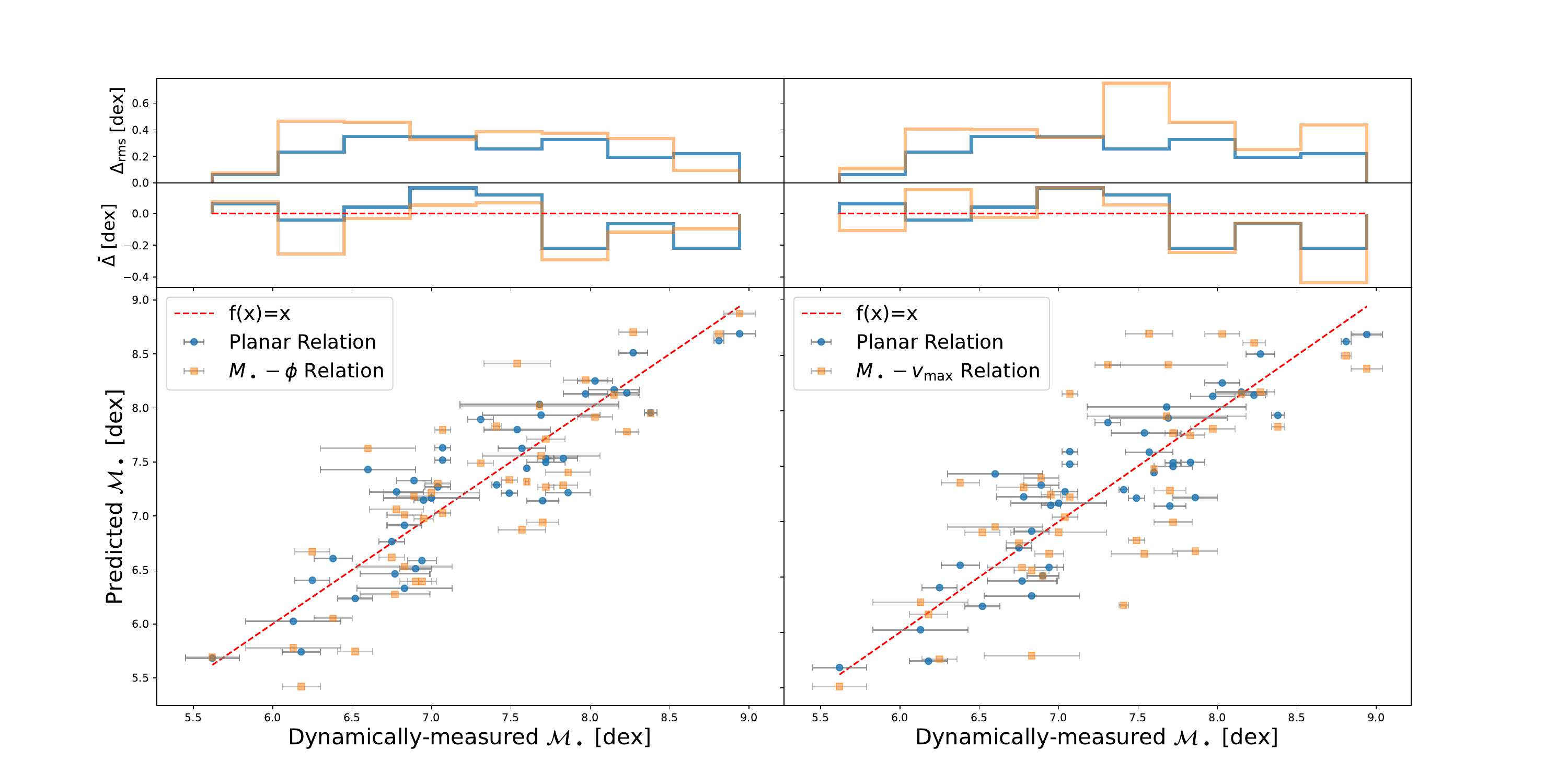}
\caption{
Plots between the dynamically-measured $M_\bullet$ on the $x$-axis and the $M_\bullet$ predicted by Equation~\ref{eqn:FP} (\textcolor{MATLAB_blue}{$\bullet$}) on the $y$-axis, compared with the $M_\bullet$--$\phi$ relation from \citet[][equation~8]{Davis:2017} in the \emph{left column} of plots and the $M_\bullet$--$v_\mathrm{max}$ relation from \citet[][equation~10]{Davis:2019} in the \emph{right column} of plots (both depicted with \textcolor{MATLAB_orange}{$\blacksquare$}).
The subplots show histograms (spread across eight bins, each 0.41\,dex wide) for the $\Delta_\mathrm{rms}$ scatter (top subplots) and $\bar{\Delta}$ average residual (bottom subplots) about the 1:1 line for both the planar relation (\textcolor{MATLAB_blue}{{\hdashrule[0.35ex]{8mm}{0.4mm}{}}}) and the $M_\bullet$--$\phi$ (\emph{left column}) or $M_\bullet$--$v_\mathrm{max}$ (\emph{right column}) relations (both depicted with \textcolor{MATLAB_orange}{{\hdashrule[0.35ex]{8mm}{0.4mm}{}}}).
The dashed line (\textcolor{red}{{\hdashrule[0.35ex]{8mm}{1pt}{1mm}}}) depicts the 1:1 correlation between observed and predicted masses in the main plots and $\bar{\Delta}=0.0$\,dex in the lower subplots.
For clarity, predicted errors (along the $y$-axis) are not shown because they are directly proportional to (and always greater than) the errors along the $x$-axis.
}
\label{fig:bicomp}
\end{figure*}

Because the $M_\bullet$--$\phi$--$v_\mathrm{max}$ relation is a combination of the $M_\bullet$--$\phi$ and $M_\bullet$--$v_\mathrm{max}$ relations (see the bottom panels of Figure~\ref{fig:FP} for projections of these relations), we begin by comparing our 3-D relation to each of the 2-D relations (see Figure~\ref{fig:bicomp}).
First (in the left column of plots in Figure~\ref{fig:bicomp}), we compare the fundamental plane with the $M_\bullet$--$\phi$ relation \citep[][equation~8]{Davis:2017}.
From a glance, we find that both relations display tight correlations without significant outliers.
For a more complete comparison, we have included subplots that break down the performance of each relation versus the planar relation across eight bins.
One subplot shows root mean square error ($\Delta_\mathrm{rms}$) in each bin and the other subplot shows the mean absolute scatter ($\bar{\Delta}$) in each bin.
As we can see, the $M_\bullet$--$\phi$--$v_\mathrm{max}$ relation generally equals or outperforms the $M_\bullet$--$\phi$ relation in all except for the most massive bin.
Although, the planar relation does tend to be biased towards slightly over-massive black holes in the middle bins, as compared to the $M_\bullet$--$\phi$ relation.

Second (in the right column of plots in Figure~\ref{fig:bicomp}), we compare the fundamental plane with the $M_\bullet$--$v_\mathrm{max}$ relation \citep[][equation~10]{Davis:2019}.
We can see that the planar relation is significantly more accurate than the $M_\bullet$--$v_\mathrm{max}$ relation in the four most massive bins.
Here, the plane is only slightly lopsided towards over-massive black hole predictions in the central bins, relative to the $M_\bullet$--$v_\mathrm{max}$ relation.
Overall, these comparisons to each of the 2-D relations demonstrates that the fundamental plane performs well, particularly so at the low-mass end, which should make it advantageous for extrapolating toward lower mass black holes, \textit{i.e.}, IMBHs.

\subsection{Explanation of the Fundamental Plane}

The existence of a tight $M_\bullet$--$\phi$--$v_\mathrm{max}$ relation built upon $M_\bullet$--$\phi$ and $M_\bullet$--$v_\mathrm{max}$ relations is not a revelation, but rather an expected consequence of numerous prior studies.
It first goes back to the so-called ``Hubble'' tuning fork diagram \citep{Jeans:1928,Hubble:1936}\footnote{For an account of the complicated provenance of the tuning fork, we suggest further reading \citep[\textit{e.g.},][]{Block:2004,Block:2015,Graham:2019b,Graham:2023}.}, which established a clear and understandable sequence that organized spiral galaxies into morphological classes based upon the prominence of their bulges and the winding geometry of their spiral arms.
To simplify morphological trends, we can use the Hubble sequence morphological stage number, $T$, where spiral galaxies are defined as $T>0$ and higher numbers are considered to be ``later'' types.
In this way, the Hubble sequence qualitatively establishes $T\propto|\phi|$ and $T\propto M_{\star,\mathrm{sph}}$, where $M_{\star,\mathrm{sph}}$ is the stellar mass of the spheroid (bulge) component of a spiral galaxy.
Decades after dissemination of the Hubble sequence, many studies conducted quantitative studies showing that indeed $|\phi|\propto T$ \citep{Kennicutt:1981,Seigar:1998,Ma:1999,Baillard:2011,Yu:2018,Diaz-Garcia:2019,Yu:2019,Yu:2020} and \citet{Davis:2019b} showed that $|\phi|\propto M_{\star,\mathrm{sph}}$.\footnote{The $\phi$--$M_{\star,\mathrm{sph}}$ relation is actually a projection of the fundamental plane of spiral structure in disk galaxies \citep{Lin:1966,Davis:2015}.}
It follows from these correlations that there should be a correlation between $\phi$ and $M_\bullet$ \citep{Seigar:2008,Berrier:2013,Davis:2017} as both are strongly correlated to bulge mass.

As for uncovering the $M_\bullet$--$v_\mathrm{max}$ relation, we can look first at the correlation between $v_\mathrm{max}$ and $T$ identified by \citet{Roberts:1978}.
By substituting $|\phi|$ as a proxy for $T$, we then arrive at the $\phi$--$v_\mathrm{max}$ relation \citep{Kennicutt:1981,Davis:2019}.
Armed with the knowledge of both $\phi$--$M_\bullet$ and $\phi$--$v_\mathrm{max}$ relations, \citet{Davis:2019} produced an $M_\bullet$--$v_\mathrm{max}$ relation that is informed by, and consistent with, the Tully-Fisher relation \citep{Tully:1977,Tiley:2019}.
Thus, we now arrive at a unified $M_\bullet$--$\phi$--$v_\mathrm{max}$ relation that is a manifestation of the gravitational potential well of a spiral galaxy.
Ergo, in more massive galaxies with deeper potential wells, we find more massive black holes, more tightly-wound spiral patterns, and higher rotational velocities.

\subsection{Fundamental Planes with Black Hole Mass}

There have been a couple prior attempts at obtaining a fundamental plane scaling relation for SMBHs.
The most relevant example is the trivariate relation between $M_\bullet$--$\sigma_e$--$R_e$, where $R_e$ is a galaxy's half-light radius and $\sigma_e$ is the stellar velocity dispersion inside an aperture equal to $R_e$ \citep{Marconi:2003,Bosch:2016}.
For the $M_\bullet$--$\sigma_e$--$R_e$ relation, \citet{Bosch:2016} found an intrinsic scatter of $\sigma=0.49\pm0.03$\,dex in the $\mathcal{M}_\bullet$-direction.
However, this is insignificant because \citet{Bosch:2016} also found an identical intrinsic scatter for the $M_\bullet$--$\sigma_e$ relation, meaning that the addition of the third parameter, $R_e$, serves no purpose, and in practice makes things worse because it introduces another variable that contributes to error propagation.
Moreover, \citet{Bosch:2016} utilized a large sample of 230 black hole mass measurements that is ``very heterogenous'' because they are derived from a variety of measurement methods, most of which are indirect and not from dynamical methods.

The other notable example is the so-called fundamental plane of black hole activity \citep{Merloni:2003,Falcke:2004}.
This plane is one between $M_\bullet$--$L_\mathrm{R}$--$L_\mathrm{X}$, where $L_\mathrm{R}$ and $L_\mathrm{X}$ are radio and X-ray luminosity, respectively.
The fundamental plane of black hole activity is based upon the interpretation of scale invariant disk--jet coupling manifesting as an empirical relation between jet power probed by radio and mass accretion rate via X-rays.
In order to perceive the intrinsic disk--jet coupling mechanism, this requires simultaneity of radio and X-ray observations to account for the duty-cycle of active galactic nuclei. 
However, given that the processes that govern the $M_\bullet$--$L_\mathrm{R}$--$L_\mathrm{X}$ relation are highly-secular, this leads to an intrinsic scatter of $\sigma=0.96\pm0.13$\,dex in the $\mathcal{M}_\bullet$-direction \citep{Kayhan:2019}, ``indicating a large amount of unexplained variance.''
Additionally, \citet{Kayhan:2022} caution against using the fundamental plane of black hole activity without additional constraints beyond just straightforward X-ray and radio observations.
All together, the noted problems with either of the aforementioned planar relations further cements the $M_\bullet$--$\phi$--$v_\mathrm{max}$ relation's superiority as a best-in-class black hole mass scaling relation for spiral galaxies.

\subsection{Implications}

One advantageous application we envision for the $M_\bullet$--$\phi$--$v_\mathrm{max}$ relation is to use it to construct black hole mass functions (BHMFs) from surveys of spiral galaxies.
Already, the $M_\bullet$--$\phi$ relation has been utilized to model the local BHMF derived from spiral galaxies \citep{Davis:2014,Fusco:2022}.
The simple addition of $v_\mathrm{max}$, which is widely available for many spiral galaxies, could better aid in modeling the shape of the BHMF with lower scatter.
This is particularly useful as the BHMF is well known at the high-mass end, but lacking clarity at the low-mass end, which is the purview of spiral galaxies.

The BHMF is virtually unknown at $M_\bullet<10^5\,\mathrm{M}_\sun$ because of a dearth of observational evidence of IMBHs.
Extrapolating the $M_\bullet$--$\phi$ \citep{Davis:2017} and $M_\bullet$--$v_\mathrm{max}$ \citep{Davis:2019} relations down to the low-mass end predicts $M_\bullet<10^5\,\mathrm{M}_\sun$ IMBHs at $|\phi|>26\fdg8\pm2\fdg3$ and $v_\mathrm{max}<130\pm15\,\mathrm{km\,s}^{-1}$.
Using these aforementioned values as inputs to Equation~\ref{eqn:FP}, we similarly find a line across the plane defining the upper-limit at $\mathcal{M}_\bullet<5.0\pm0.4$\,dex.
Thus, these values of $|\phi|$ and $v_\mathrm{max}$ serve as a sort of midline path down the plane into the IMBH regime.
However, because of the flexibility of the plane, these need not be hard and fast values for identifying potential IMBH-hosting galaxies.
That is, a galaxy with a slightly smaller $|\phi|$ in conjunction with a slightly larger $v_\mathrm{max}$, or \textit{vice versa}, could still lie below $M_\bullet=10^5\,\mathrm{M}_\sun$ on the fundamental plane.
Particularly, in a forthcoming work \citep{Davis:2023b}, we use the fundamental plane to identify strong candidates for IMBH hosts among a sample of late-type spiral galaxies.

\section{Conclusions}\label{sec:conclusions}

Arguably, spiral galaxies are the most interesting galaxies.
This is not just because of their intrinsic beauty, but also because they are galactic laboratories of ongoing star formation, growth, and evolution.
Moreover, the most interesting discoveries await in the realm of low-mass spiral galaxies as potential hosts of elusive IMBHs.
However, these interesting characteristics make them more difficult to analyze than their more massive, simpler, and older cousins, \textit{i.e.}, early-type galaxies.
Indeed, commonly-used black hole mass scaling relations like the $M_\bullet$--$\sigma_0$ (central/bulge stellar velocity dispersion) or $M_\bullet$--$M_{\star,\mathrm{sph}}$ relations are far less accurate for late-type galaxies than early-type galaxies.
Specifically, the $M_\bullet$--$\sigma_0$ relation has an intrinsic scatter of $\sigma=0.32$\,dex for early-type galaxies \citep{Sahu:2019}, \textit{cf.}\ $\sigma=0.57$\,dex for late-type galaxies \citep{Davis:2017,Sahu:2019}; the $M_\bullet$--$M_{\star,\mathrm{sph}}$ relation has an intrinsic scatter of $\sigma=0.41$\,dex for early-type galaxies \citep{Sahu:2019b}, \textit{cf.}\ $\sigma=0.48$\,dex for late-type galaxies \citep{Davis:2019b,Sahu:2019b}.
Judged upon these criteria, the $M_\bullet$--$\phi$--$v_\mathrm{max}$ relation (with $\sigma=0.22\pm0.06$\,dex) is not only more accurate than other late-type black hole mass scaling relations, but also those for early-type galaxies.

Verily, one might expect that a tighter relationship would exist for single-component galaxies like elliptical galaxies, rather than multi-component spiral galaxies.
However, if we consider the evolution of these systems and the increase in entropy from spiral to elliptical galaxies, it becomes evident that this initial assumption may not hold.
Although late-type galaxies may be \emph{complex}, early-type galaxies are \emph{complicated}; the distinction being that \textbf{complexity} implies \underline{many understandable components} and \textbf{complication} implying less components, but \underline{more chaos and disorder}.
This can be understood by tracking the impact of merger histories and the genesis of morphologically-dependent black hole mass scaling relations \citep{Graham:2023a,Graham:2023b,Graham:2023c,Graham:2023d}.
As such, the unique merger history of a galaxy can effectively muddle the ordered rotationally-supported disk galaxies by transforming them into dispersion-supported elliptical galaxies.
Moreover, this helps explain why we find that the strongest correlation with black hole mass is not via bulge properties, which are similar to the disordered spheroids of elliptical galaxies and may be the result of disk cloaking \citep{Hon:2022}.

The fact that the $M_\bullet$--$\phi$--$v_\mathrm{max}$ relation, which correlates the black hole mass with global properties of its host galaxy's disk rather than bulge properties, as in the $M_\bullet$--$\sigma_0$ and $M_\bullet$--$M_{\star,\mathrm{sph}}$ relations, shows that BH--galaxy coevolution is active over large scales.
Indeed, recent work \citep{Davies:2019,Oppenheimer:2020,Sanchez:2023} has shown that $M_\bullet$ is inversely correlated with the fraction of baryons in the circumgalactic medium of its host galaxy.
This is thought to be because more massive black holes are more energetic, and thus transport more baryons beyond their host galaxies' virial radii, all while reducing gas accretion and star formation over time.
This is clear evidence that processes of a central SMBH are capable of affecting change on scales over eleven orders of magnitude larger than extent of their event horizons!\footnote{The Milky Way's Sgr~A$^\ast$ has a shadow with a radius of $0.21\pm0.01$\,AU \citep{Akiyama:2022} and the virial radius of the Galaxy is 258\,kpc \citep{Klypin:2002}, which is an astounding difference in scale of $(2.52\pm0.12)\times10^{11}$.
For comparison, there is similar difference in scale between the width of a human hair and the radius of Earth.}

Extrapolation of black hole mass scaling relations down into the IMBH range is important for future studies, including the design and predictions for space-based gravitational-wave interferometers \citep{LISA:2022}.
Therefore, we anticipate that the fundamental plane will be advantageous for estimating the demographics of IMBHs hosted by spiral galaxies.
With more than one parameter, there is redundancy built into the planar relationship that makes it more resilient to abnormalities in a single parameter, helping it to be more robust.
This adds a degree of confidence when using it to predict black holes masses below the limits of our sample (NGC~4395 with $\mathcal{M}_\bullet=5.62\pm0.17$\,dex).
Moreover, spiral-arm pitch angle is straightforward enough to measure that it could be accomplished most basically with just an uncalibrated image and a protractor.
What is more, $v_\mathrm{max}$ values are readily-available from large 21-cm line width surveys and easily-accessible in online archives.
Therefore, we hope that the $M_\bullet$--$\phi$--$v_\mathrm{max}$ relation will facilitate new and impactful studies and influence further advancements in black hole mass scaling relations and galaxy evolution.

\begin{acknowledgments}
The authors are grateful for stimulating discussions with Andrea Macci\`{o}, Joseph Gelfand, Ingyin Zaw, and Ivan Katkov.
This material is based upon work supported by Tamkeen under the NYU Abu Dhabi Research Institute grant CASS.
This research has made use of NASA's Astrophysics Data System, and the NASA/IPAC Extragalactic Database (NED) and Infrared Science Archive (IRSA).
We acknowledge the use of the HyperLeda database (\url{http://leda.univ-lyon1.fr}).
\end{acknowledgments}

\software{
\\
\href{https://github.com/astropy/astropy}{\textcolor{linkcolor}{\texttt{Astropy}}} \citep{astropy:2013,astropy:2018}\\
\href{https://github.com/CullanHowlett/HyperFit}{\textcolor{linkcolor}{\texttt{Hyper-Fit}}} \citep{Robotham:2015,Robotham:2016}\\
\href{https://github.com/matplotlib/matplotlib}{\textcolor{linkcolor}{\texttt{Matplotlib}}} \citep{Hunter:2007}\\
\href{https://github.com/numpy/numpy}{\textcolor{linkcolor}{\texttt{NumPy}}} \citep{harris2020array}\\
\href{https://pandas.pydata.org/}{\textcolor{linkcolor}{\texttt{Pandas}}} \citep{McKinney_2010}\\
\href{https://github.com/MilesCranmer/PySR/tree/v0.12.3}{\textcolor{linkcolor}{\texttt{PySR}}} \citep{Cranmer:2023}\\
\href{https://www.python.org/}{\textcolor{linkcolor}{\texttt{Python}}} \citep{Python}\\
\href{https://github.com/scipy/scipy}{\textcolor{linkcolor}{\texttt{SciPy}}} \citep{Virtanen_2020}\\
\href{http://pythonhosted.org/uncertainties/}{\textcolor{linkcolor}{\texttt{uncertainties}}}
}

\section*{ORCID iDs}

\begin{CJK*}{UTF8}{gbsn}
\begin{flushleft}
Benjamin L.\ Davis \scalerel*{\includegraphics{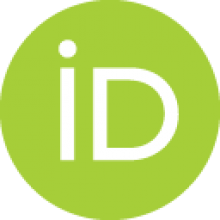}}{B} \url{https://orcid.org/0000-0002-4306-5950}\\
Zehao Jin (金泽灏) \scalerel*{\includegraphics{orcid-ID.png}}{B}\\\url{https://orcid.org/0009-0000-2506-6645}
\end{flushleft}
\end{CJK*}

\appendix

\section{The Milky Way}\label{app:MW}

\begin{figure*}
\includegraphics[clip=true, trim= 31mm 11mm 49mm 25mm, width=\textwidth]{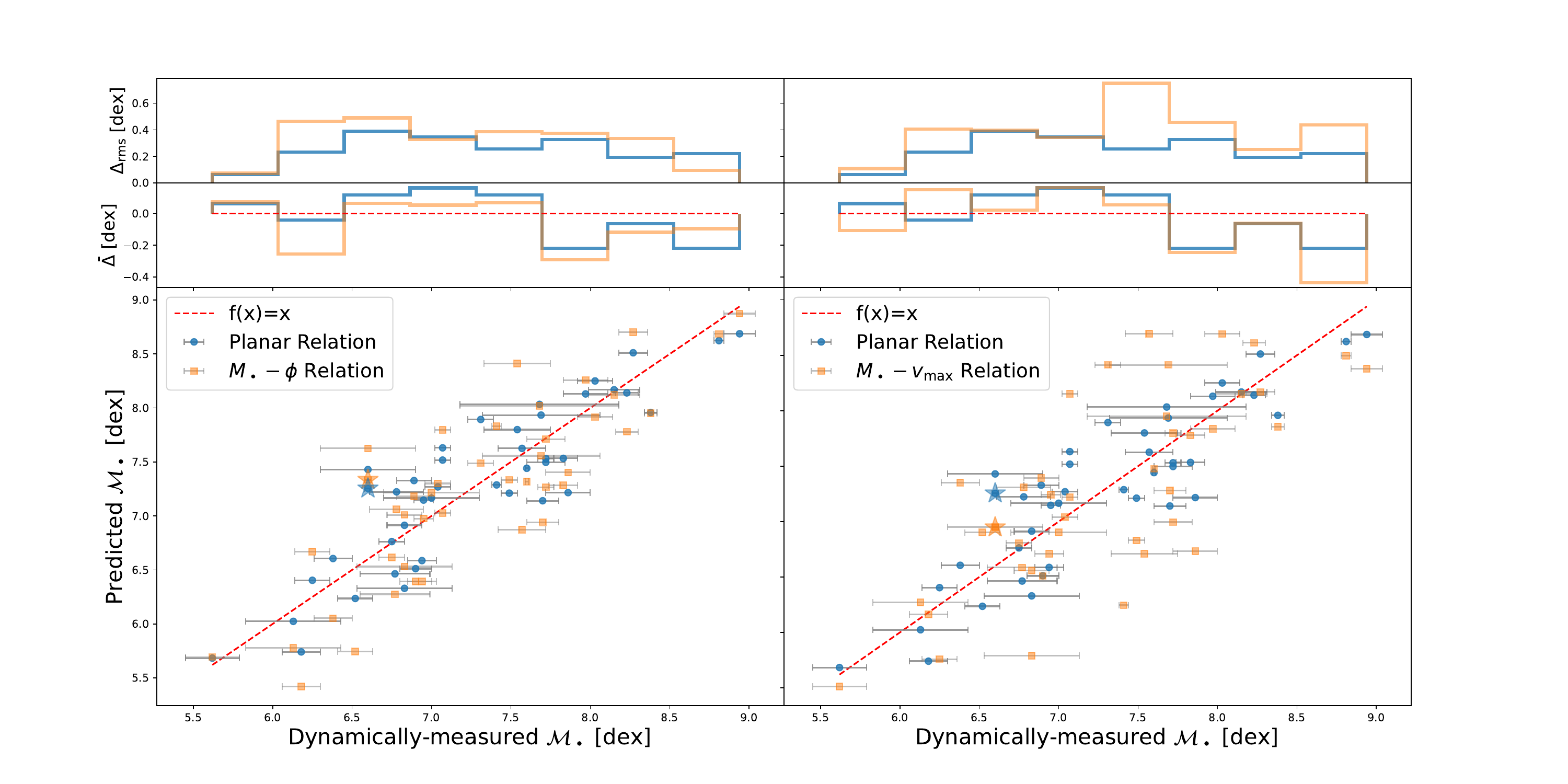}
\caption{
This figure is identical to Figure~\ref{fig:bicomp}, except it includes the Milky Way (marked with $\star$).
For an animation of Figure~\ref{fig:FP} that also includes the Milky Way, see the following link, \url{http://surl.li/iggks}.
}
\label{fig:MW}
\end{figure*}

Sgr~A$^\ast$ in our Galaxy has been robustly studied by many independent methods, determining its black hole mass with incredible accuracy and precision.
We adopt the mass determined by the multi-star orbit analysis of \citet{Boehle:2016}, but many such other studies present consistent masses, most notably the black hole mass determined by the size of its shadow \citep{EHT:2022}.
Therefore, we strongly intended to include the Milky Way in our sample, just as it was included in the determination of the $M_\bullet$--$\phi$ and $M_\bullet$--$v_\mathrm{max}$ relations.
However, our home galaxy stands out as a significant outlier below the $M_\bullet$--$\phi$--$v_\mathrm{max}$ plane, \textit{i.e.}, its dynamically-measured $M_\bullet$ is under-massive with respect to that predicted by the plane.
Specifically, the fundamental plane predicts $\mathcal{M}_\bullet=7.26\pm0.24$\,dex, whereas Sgr~A$^\ast$ is highly-constrained to $\mathcal{M}_\bullet=6.60\pm0.02$\,dex.
You can see how the Milky Way stands out in Figure~\ref{fig:MW}.

In absolute terms, this does not necessarily make the Milky Way the most extreme outlier in the sample, but the accuracy of its dynamically-measured $M_\bullet$ means that it is weighted heavily in the regression of the fundamental plane. 
The coefficients for Equation~\ref{eqn:FP}, with the Milky Way included, changes to $\alpha=-5.57\pm0.06$, $\beta=3.95\pm0.06$, $\gamma=7.31\pm0.06$, and intrinsic scatter $\sigma=0.28\pm0.06$\,dex in the $\mathcal{M}_\bullet$-direction.
This represents a small change in the predicted black hole mass; for a galaxy with the median $\phi$ and $v_\mathrm{max}$, the plane without the Milky Way yields $M_\bullet=(2.16\pm1.13)\times10^7\,\mathrm{M}_\sun$ and including the Milky Way it becomes $M_\bullet=(2.06\pm1.36)\times10^7\,\mathrm{M}_\sun$.
Thanks to its position relatively near the balance point of the fundamental plane, its affect does not noticeably tug the plane off in any direction.
However, it does diminish the accuracy of the fundamental plane, as evidenced by the 0.06\,dex increase in the intrinsic scatter due entirely to our Galaxy.
Thus, one can choose to use the alternative coefficients for Equation~\ref{eqn:FP} that considered the Milky Way and yield highly-consistent black hole mass predictions with only a small decrease in accuracy, which is still more accurate than either the $M_\bullet$--$\phi$ or $M_\bullet$--$v_\mathrm{max}$ relations alone.

Since the mass of Sgr~A$^\ast$ is practically unassailable, the fault in our Galaxy must lie in its $\phi$ and/or $v_\mathrm{max}$.
Of course, both quantities are difficult to measure from inside the Galaxy; even with modern data, it is hard to truly represent the geometric shape that astronomers from the Andromeda Galaxy would see or what $v_\mathrm{max}$ they would measure from long-slit spectroscopic observations of the Milky Way.
As shown in Figure~\ref{fig:MW}, the $M_\bullet$--$\phi$ relation predicts a less accurate black hole mass, whereas the $M_\bullet$--$v_\mathrm{max}$ relation actually predicts a more accurate black hole mass than the fundamental plane.
Indeed, rearranging Equation~\ref{eqn:FP} to solve for $\phi$ yields a prediction of $|\phi|=19\fdg3\pm2\fdg1$ for the Galaxy, which is not far-fetched to envision.
Our adopted value of $13\fdg1\pm0\fdg6$ \citep{Jacques:2015} is the median pitch angle derived from a meta-analysis of 50 studies with a range of $3\degr\leq|\phi|\leq28\degr$.
Moreover, with our intimate vantage of the Milky Way, observations can be overwhelmed with small-scale structures, such as a high pitch angle structure in the Sagittarius Arm \citep{Kuhn:2021}, that can complicate determinations of the global Galactic pitch angle.

For a final consideration, it could be that Sgr~A$^\ast$ is simply under-massive.
Indeed, \citet{Oppenheimer:2020} point out that Sgr~A$^\ast$ is under-massive with respect to other SMBHs in galaxies with similarly-sized halos as the Milky Way.
In fact, it is thought that an under-massive Sgr~A$^\ast$ could be conducive to supporting the genesis of life in the Milky Way \citep[\textit{e.g.},][]{Lingam:2019}.
Thus, the Milky Way, and its central black hole, must be at least consistent with the anthropic principle \citep{Dicke:1957}.

\bibliography{bibliography}

\end{document}